\documentclass[11pt,a4paper]{article}
\usepackage{jheppub}
\usepackage{color,subfigure}

\title{$q-$form fields on $p-$branes}
\author{Chun-E Fu,
        Yu-Xiao Liu\footnote{Corresponding author},
        Ke Yang,
        Shao-Wen Wei}
\affiliation[]{Institute of Theoretical Physics,
    Lanzhou University, Lanzhou 730000,
    People's Republic of China}
\emailAdd{fuche08@lzu.edu.cn}
\emailAdd{liuyx@lzu.edu.cn}
\emailAdd{yangke09@lzu.edu.cn}
\emailAdd{weishw@lzu.edu.cn}

\abstract{In this paper, we give one general method for localizing any form ($q-$form) field on $p-$branes with one extra dimension, and apply it to some typical $p-$brane models. It is found that, for the thin and thick Minkowski branes with an infinite extra dimension, the zero mode of the $q-$form fields with $q<(p-1)/2$ can be localized on the branes. For the thick Minkowski $p-$branes with one finite extra dimension, the localizable $q-$form fields are those with $q<p/2$, and there are also some massive bound Kaluza-Klein modes for these $q-$form fields on the branes. For the same $q-$form field, the number of the bound Kaluza-Klein modes (but except the scalar field ($q=0$)) increases with the dimension of the $p-$branes. Moreover, on the same $p-$brane, the $q-$form fields with higher $q$ have less number of massive bound KK modes.  While for a family of pure geometrical thick $p-$branes with a compact extra dimension, the $q-$form fields with $q<p/2$ always have a localized zero mode.
For a special pure geometrical thick $p-$brane, there also exist some massive bound KK modes of the $q-$form fields with $q<p/2$, whose number increases with the dimension of the $p-$brane.
}

\begin{document}
\maketitle

\section{Introduction}

The idea of extra dimension was proposed to unify different forces. The representative one is the Kaluza-Klein (KK) theory, which unifies the 4-dimensional (4D) gravity and electromagnetism by a 5-dimensional Einstein's theory with a spatial extra dimension. And the string/M-theory also requires that there are 6/7 extra dimensions. But the extra dimensions in these theories are so extremely small that can not be detected by the present experiments.

When the Arkani-Hamed-Dimopoulos-Dvali (ADD) \cite{Antoniadis1998} and the Randall-Sundrum (RS) \cite{Randall1999a,Randall1999b} brane-world models were put forward, the extra dimension theory was revived. One of the reasons is because the scale of extra dimensions can be millimetre or even infinite, which increases the possibility to be detected in experiment. Another one is because the theory opens a new way to solve the long-standing hierarchy problem and the cosmology problem \cite{ArkaniHamed1998rs,Cosmological2000,coscon2009,Cosmoligicalbrane2010}.

{In the Randall-Sundrum model, the Standard Model (SM) fields are confined on the brane, and gravity propagates in the whole space-time. This idea is motivated by the string theory, in which the SM fields are excitation modes of open string whose ends lie on the brane, and graviton is a closed string excitation having no-end points, thus it does not have to be attached to the brane. Moreover, the $q-$form fields also comprise the excitation of a closed string \cite{stringvol.1,stringvol.2}. So it is reasonable to introduce the $q-$form fields in the RS (or RS-like) brane, and see the localization of these fields on different branes, including the thin and thick branes \cite{Rubakov1,Rubakov2,ThickBraneDewolfe,Thickbrane2000,
ThickBrane2001,ThickBrane2002,ThickBrane2003,Blochbrane2004,
ThickBraneBazeia2006,ThickBrane20071,ThickBrane20082,ThickBrane20093,
KeYang2009Weyl,ThickBraneZhongYuan2011JHEP,ThickBraneKeYang2011PRD,
ThickBraneCriticalGravityLiu2012,BraneFR2012}.

On the other hand, although free q-form fields in the bulk are just equivalent to scalar or vector fields by a duality in 4D space-time, they are new types of particles in the higher space-time.

In this work, we consider the branes with $p$ spacial dimensions in the $D=p+2$ dimensional space-time, which are called $p-$branes as in the string/M-theory. The realistic world is the $3-$brane, and the higher dimensional branes with $p\geq5$ ($D\geq7$) may also have realistic applications if the branes have 3 infinite large dimensions (which are those we can feel) and $p-2$ finite size dimensions with topology $S^1 \times S^1 \times ... \times S^1 = T^{p-2}$ and small enough radius. Thus we will investigate the localization of $q-$form fields on these $p-$branes.}

The introduction of extra dimensions implies that there are KK modes accompanying all the particles on the brane \cite{WarpedPassages}. These KK modes carry the messages of the extra dimensions. If the KK modes can be localized on the brane, there is hope to detect them and prove the existence of the extra dimensions, which is one main meaning of  localization. The massless KK modes exactly stand for the particles on the brane, and surely must be localized for a realistic brane-world model. And the next heavier massive modes are the key clue.

In Refs.~\cite{LocalizationPRDLanglois2003,
LocalizationGravityJHEP2005,
Liu2007WeylVo,Liu2008WeylPT,LocalizationMO2008,JHEPadsGuo2009,
LocalizationFermionSplit2010,
LocalizationFuPRD2011,PRDefromedLiu,LocalizationCastro2011,
LocalizationZhao2011JHEP,LocalizationWithoutScalar2011JHEP,
LocalizationWaveBrane2012,LocalizationFermion2012}, the localization of the scalar, vector and fermion fields has been investigated. In this paper, we try to find one general method for localizing any form ($q-$form) field. For $q=0,1$ the fields are just the scalar and vector ones, and for higher $q$ they stand for new particles in the higher space-time \cite{Quevedo2010ui-eprintv1}. In the RS model, the localization of $2-$form and $3-$form fields has been investigated \cite{PRLKR2002,Mukhopadhyaya2004,QformRS,qRSdilation-eprintv2}, and it was shown that it is possible to find some signal about the fields coupled with a dilaton at Tev scale. Then the authors discussed the localization of the $q-$form fields on a domain wall in a RS-like scenario \cite{qdual2010-eprintv3}.

While our aim is to find the general method for localizing the $q-$form fields, and then apply this to some typical branes. In the following, we first study how to obtain the effective potential of KK modes for the $q-$form fields in Sec. \ref{SecLocalize}. Then, with some typical $p-$brane models in Sec. \ref{pbranes}, we investigate the condition for localizing the $q-$form fields and the effect of the dimension of the $p-$branes. Finally, we give a simple conclusion in Sec. \ref{conclusion}.

\section{How to localize $q-$form fields}
\label{SecLocalize}

We consider a free antisymmetric $q-$form field $X_{M_1M_2...M_q}$ on $p-$branes in $D-$dimensional space-time with $D=p+2$, and treat the field as a small perturbation around the background. The action for the field is
\begin{eqnarray}\label{qaction}
S_q=\int d^D x\sqrt{-g}\;Y_{M_1M_2... M_{q+1}}Y^{M_1M_2... M_{q+1}},
\end{eqnarray}
where $Y_{M_1M_2...M_{q+1}}$ is the field strength defined as $Y_{M_1M_2...M_{q+1}}=\partial_{[M_1}X_{M_2...M_{q+1}]}$.
Because the $(D-1)-$form and higher form fields have no role to play in the brane \cite{QformRS}, we only analyze the cases $q=0,1,2,...,p$.

The line-element of the $D-$dimensional space-time is assumed as
\begin{eqnarray}\label{line-element}
 ds^2&=&\text{e}^{2A(z)}\big(\eta_{\mu\nu}dx^\mu dx^\nu
          + dz^2\big)
\end{eqnarray}
with $z$ denoting the extra dimension vertical to the brane. Then the equations of motion can be derived from the action (\ref{qaction}) and the conformal metric (\ref{line-element}):
\begin{eqnarray}
 \partial_{\mu_1} ( \sqrt{-g}\;Y^{\mu_1\mu_2...\mu_{q+1}})
 +\partial_z(\sqrt{-g}\;Y^{z\mu_2...\mu_{q+1}})&=& 0, \label{equ1}\\
 \partial_{\mu_1} ( \sqrt{-g}\;Y^{\mu_1...\mu_qz})&=& 0.
\end{eqnarray}

In order to investigate the localization of the $q-$form field on a $p-$brane, we first choose a gauge $X_{\mu_1...\mu_{q-1}z}=0$ using the gauge free $\delta X_{M_1M_2...M_q}=\partial_{[M_1}\Lambda_{M_2...M_q]}$ with $\Lambda_{M_2M_3...M_q}$ an antisymmetric tensor \cite{Mukhopadhyaya:2007jn}, and make a decomposition for the $X_{\mu_1\mu_2...\mu_q}$:
\begin{eqnarray}\label{KK1}
X_{\mu_1\mu_2...\mu_q}(x^\mu,z)=\sum_n
\hat{X}^{(n)}_{\mu_1\mu_2...\mu_q}(x^\mu)U^{(n)}(z) \text{e}^{(2q-p)A/2},
\end{eqnarray}
so the field strength becomes
\begin{eqnarray}
Y_{\mu_1\mu_2...\mu_{q+1}}(x^\mu,z)&=&\sum_n \hat{Y}^{(n)}_{\mu_1\mu_2...\mu_{q+1}}(x^\mu)
U^{(n)}(z) \text{e}^{(2q-p)A/2},\label{KK10}\\
Y_{\mu_1\mu_2...\mu_qz}(x^\mu,z)&=&\sum_n\frac{1}{q+1}\hat{X}^{(n)}_{\mu_1\mu_2...\mu_q}(x^\mu)
\left(U^{(n)}{'}+\frac{2q-p}{2}A'U^{(n)}\right)U^{(n)}(z) \text{e}^{(2q-p)A/2},~~~~~\label{KK11}
\end{eqnarray}
where $\hat{Y}_{\mu_1...\mu_{q+1}}(x^\mu)=\partial_{[\mu_1}\hat{X}_{\mu_2...\mu_{q+1}]}(x^\mu)$ is the field strength on the brane.

Substituting the relations (\ref{KK10}) and (\ref{KK11}) into the equations of motion (\ref{equ1}), we can get the following Schr\"{o}dinger-like equation for the KK modes of the $q-$form field:
 \begin{eqnarray}
\big[ -\partial^2_z+ V(z)\big]U^{(n)}(z)
  =m_n^2 U^{(n)}(z),
  \label{SchEq}
\end{eqnarray}
where $m_n$ are the masses of the KK modes, and the effective potential $V(z)$ takes the form
\begin{eqnarray}\label{V}
V(z)=\frac{(p-2q)^2}{4}A'^2(z) +\frac{p-2q}{2}A''(z)
\label{Veff}
\end{eqnarray}
with the prime standing for the derivative with respect to $z$.

Furthermore, provided the orthonormality condition
\begin{equation}\label{NormCondi}
  \int U^{(m)}U^{(n)}dz=\delta_{mn},
\end{equation}
the effective action of the $q-$form field on the brane could be obtained:
\begin{eqnarray}
S_{\text{eff}} = \sum_n \int d^{D-1}x\sqrt{-\hat{g}}
\bigg(\hat{Y}^{(n)}_{\mu_1\mu_2...\mu_{q+1}}\hat{Y}^{(n)\mu_1\mu_2...\mu_{q+1}}
+\frac{1}{q+1}m_n^2\hat{X}^{(n)}_{\mu_1\mu_2...\mu_q}\hat{X}^{{(n)}\mu_1\mu_2...\mu_q}
\bigg).
\end{eqnarray}

For $m_0^2=0$, we get the zero mode for the $q-$form field from (\ref{SchEq}):
\begin{equation}\label{U0}
  U_0\propto \text{e}^{\frac{p-2q}{2}A},
\end{equation}
which results in $X^{(0)}_{\mu_1\mu_2...\mu_q}(x^\mu,z)=
\hat{X}^{(0)}_{\mu_1\mu_2...\mu_q}(x^\mu)$ (see from (\ref{KK1})), and it is irrelevant to the extra dimension.
It means that the KK mode with $m^2=0$ (the zero mode) exactly reproduces the particle on the brane. It surely should be localized on the realistic branes. And we can judge this with the help of the  orthonormality condition (\ref{NormCondi}). So the messages of the extra dimension only can be disclosed by the massive KK modes. We could investigate the feature of them using the Schr\"{o}dinger-like equation (\ref{SchEq}). At last the core of the problem is focused on the effective potential in (\ref{SchEq}), which is decided by three factors, i.e., the parameters $q$, $p$, and the background space-time. And this is the clue of the following work.

\section{Localization of $q-$form fields on different $p-$branes}
\label{pbranes}

In the section we try to investigate the localization of the $q-$form fields on different types of flat $p-$branes, i.e., the thin RS $p-$branes, the thick ones with finite and infinite extra dimension, and the pure geometrical ones. So we first give the solutions of the $p-$branes. The effective potentials on these branes mainly have two types: the volcano-like type and the P\"{o}schl-Teller (PT)-like one. For the former type there exist only continuous massive KK modes, but for the later one some massive bound KK modes appear. The effect of the parameters $p$ and $q$ on the localization will also be discussed.

\subsection{Thin $p-$brane model}

Firstly, we consider thin $p-$brane (RS brane) in $D-$dimensional AdS space-time with a negative cosmological constant $\Lambda$. The action of this system is
\begin{eqnarray}
  S=\frac{1}{2\kappa_D^2}\int d^Dx\sqrt{-g}\big[R-p\;\Lambda\big]
        + \int d^{D-1}x \sqrt{-g^{(b)}}(-\sigma),
\end{eqnarray}
where $g^{(b)}_{\mu\nu}$ is the induced metric on the brane, $\sigma$ is the brane tension, and $\kappa_D^2=8 \pi G_D$ with $G_D$ the D-dimensional Newton constant. Here we let $\kappa_D=1$. Then the Einstein equation reads:
\begin{equation}
  G_{MN}+\frac{\Lambda}{2}p\;g_{MN}
  =-\sigma\delta^\mu_M\delta^\nu_N\eta_{\mu\nu}\text{e}^A\delta(z),
\end{equation}
where $G_{MN}=R_{MN}+\frac{1}{2}g_{MN}R$ is the Einstein tensor. Then the solution of the thin flat $p-$brane is
\begin{eqnarray}
  A(z)&=&-\ln{\big(1+\sqrt{\frac{-\Lambda}{p+1}}|z|\big)},\label{WarpRS}\\
  \sigma&=&2\;p\sqrt{\frac{-\Lambda}{p+1}}.
\end{eqnarray}

So in this brane model, the form of the potential (\ref{V}) is expressed as
\begin{eqnarray}
  V(z)=-\frac{(p-2q)(p-2q+2)\Lambda}{p+1}\frac{1}{4(1+k_1|z|)^2}
       -\frac{k_1(p-2q)}{1+k_1|z|}\delta(z)
\end{eqnarray}
with $k_1=\sqrt{\frac{-\Lambda}{p+1}}$. As $\Lambda$ is negative, the potential is volcano-like for $q<p/2$. And there is a zero mode for the $q-$form field with the wave function (\ref{U0}). We can check whether it can be localized on the brane through (\ref{NormCondi}):
\begin{eqnarray}
  \int U_0^2dz=\int (1+k_1 |z|)^{2q-p}dz=\frac{2}{k_1(2q-p+1)}(1+k_1 |z|)^{2q-p+1},
\end{eqnarray}
from which it is clear that, for $q<\frac{p-1}{2}$ the integral is finite, hence, the zero mode can be localized on the brane. For example, for $p=3$, only the scalar field $(q=0)$ can be localized. While for higher dimension case, the vector field can be as well. { And the $2-$form field can also not be localized on the $3-$brane, such as the $2-$form Kalb-Ramond (KR) field. This result corresponds to that in Refs.~\cite{PRLKR2002,Mukhopadhyaya2004,LocalizationFuPRD2011,PRDefromedLiu}. It is shown that the KR field can be considered as a source of torsion \cite{PRLKR2002,Mukhopadhyaya2004}, however, there is no phenomenon about the torsion of our space-time ($3-$brane). Here, we can give a possible reason for this, i.e., the KR field should be localized on the higher dimensional brane ($p\geq5$). And for the following brane model, we will obtain the same result for the KR field.}

\subsection{Thick flat $p-$branes model}

In this subsection, we will investigate the $q-$form field on thick Minkowski $p-$branes, which are generated by one or two scalar fields.

\subsubsection{Thick flat $p-$brane with infinite extra dimension}

Considering the thick $p-$brane generated by one scalar field $\phi$, the action is given by
\begin{equation}
S=\int d^Dx \sqrt{-g}
 \big[\frac{1}{2\kappa_D^2}R -\frac{1}{2}(\partial\phi)^2-V(\phi)\big].
\end{equation}
 With the metric
\begin{equation}\label{line-element-y}
ds^2=\text{e}^{2A}\eta_{\mu\nu}dx^\mu dx^\nu+dy^2
\end{equation}
the equations of motion can be get:
\begin{eqnarray}
  \phi'^2&=&-p\;A'',\\
  -2V&=&p\;(p+1)\;A'^2+p A'',\\
  dV/d\phi&=&\phi''+(p+1)A'\phi',
\end{eqnarray}
where the prime denotes the derivative with respect to $y$. Then we use the superpotential method to solve these equations. With the introduction of the superpotential function $W(\phi)$ defined as $\phi'=dW/d\phi$, the above second-order differential equations are reduced to the following one-order differential ones:
\begin{eqnarray}
  A'=-\frac{1}{p}W,~~~
  V=-\frac{1}{2}\left[\frac{p+1}{p}W^2-(\frac{d W}{d\phi})^2\right].
\end{eqnarray}
If we suppose that the superpotential is the sine-Gordon type, i.e., $W(\phi)=\frac{3}{2}bc\sin{\sqrt{\frac{2}{3b}}\phi}$ with $b$ and $c$ both constants, the solution of this flat thick brane can be obtained:
\begin{eqnarray}
  \phi(y)&=&\sqrt{6b}\arctan{\tanh{(cy/2)}},\\
  A(y)&=&\frac{3b}{2p}\ln{\text{sech}(cy)}\label{Solu-sineGorden}.
\end{eqnarray}

With the coordinate transformation $dz=\text{e}^{-A}dy$, the warp factor tends to $A(z)\rightarrow-\ln{(\frac{3bc}{2p}z+1)}$ when $z \rightarrow \pm\infty$. Then the behavior of the potential $V(z)$ at $z=0$ and at infinity is
\begin{eqnarray}
  V(z=0)&\rightarrow&-\frac{3b^2c}{4p}(p-2q),\\
  V(z\rightarrow\pm\infty)&\rightarrow&0.
\end{eqnarray}
For $q<p/2$, the potential is also a volcano-like one. So there is a zero mode for the $q-$form field. With the orthonormality condition (\ref{NormCondi})
\begin{eqnarray}
  \int U_0^2dz\rightarrow\int(1+\frac{3bc}{2p}z)^{-(p-2q)}dz,
\end{eqnarray}
we see that, for $q<\frac{p-1}{2}$, the zero mode can be localized on the brane, which is the same with the thin brane case.

\subsubsection{Thick flat $p-$brane with finite extra dimension}
\label{flatfinite}

Now let's turn to another kind of thick flat $p-$brane, which is generalized by two scalar fields $\phi$ and $\pi$. The action of this system is
\begin{eqnarray}
  S=\int d^D x \sqrt{-g}\bigg[\frac{1}{2\kappa_D^2}R-\frac{1}{2}\phi'^2-\frac{1}{2}\pi'^2
  -V(\phi,\pi)\bigg].
\end{eqnarray}
Then, with the conformally flat metric (\ref{line-element}), the equations of motion read
\begin{eqnarray}
  \phi'^2+\pi'^2&=&p(A'^2-A''),\\
  -2\text{e}^{2A}V&=&p^2A'^2+p A'',\\
  dV/d\phi&=&\text{e}^{-2A}\big[\phi''+p A'\phi'\big],\\
  dV/d\pi&=&\text{e}^{-2A}\big[\pi''+p A'\pi'\big].
\end{eqnarray}
Using the superpotential method, we can find that the following one-order differential equations are the solutions of the above ones:
\begin{eqnarray}
  A'=-\frac{1}{p}W, ~~~~\pi=\sqrt{p}\;A,~~~~
  V=-\frac{1}{2}\text{e}^{-\frac{2}{\sqrt{p}}\pi}
  \left[W^2-(\frac{\partial W}{\partial \phi})^2\right].
\end{eqnarray}
Supposing the form of the superpotential as $W(\phi)=va\phi(1-\frac{\phi^2}{3v^2})$ with $v$ and $a$ positive constants, we obtain the following solution:
\begin{eqnarray}
  \phi&=&v \tanh{(az)},\\
  A&=&-\frac{v^2}{3p}\left[\ln{\cosh^2{(az)}}+\frac{1}{2}\tanh^2{(az)}\right],\\
  \pi&=&\sqrt{p}\;A.
\end{eqnarray}
For this solution, the physical extra dimension $y$ is finite, which can be seen from the coordinate transformation $\int dy=\int \text{e}^{A}dz\propto \int \text{e}^{-k_2z}dz$ with $k_2$ a positive constant. And we have discussed the finity of the extra dimension in Ref.~\cite{PRDefromedLiu} in 5D case.

\begin{figure*}[htb]
\begin{center}
\includegraphics[width=6cm]{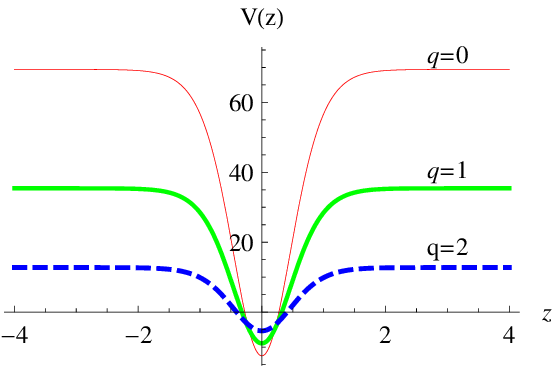}
~~~~\includegraphics[width=6cm]{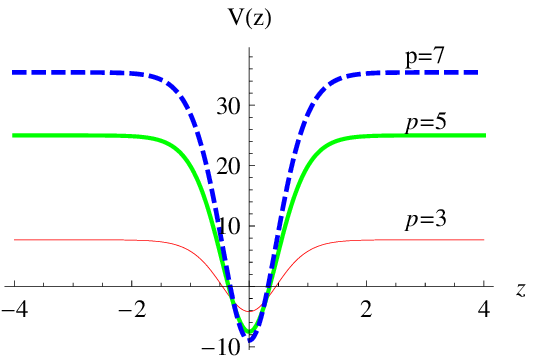}
\end{center}
 \caption{The shapes of the effective potential $V(z)$ (\ref{VzFiniteED}) for $v=5,a=1$. The left one is for different $q$ with $p=7$, and the right one is for different $p$ with $q=1$.}
 \label{FigVfinite}
\end{figure*}

For this brane-world model, the effective potential $V(z)$ reads
\begin{eqnarray}
  V(z)=\frac{a^2v^2(p -2 q)}{36p^2}
   \bigg[-18 p\;\text{sech}^4{(az)} + (p-2q)v^2\tanh^2{(az)} \big[2 + \text{sech}^2(az)\big]^2 \bigg],
   \label{VzFiniteED}
\end{eqnarray}
whose asymptotic behavior is
\begin{eqnarray}
 V(z=0)&=&-\frac{a^2 v^2(p - 2 q)}{2 p},\\ V(z\rightarrow\infty)&\rightarrow&\frac{a^2v^4(p - 2 q)^2}{9 p^2}.
\end{eqnarray}
For $q<p/2$, the potential is a PT-like one, and there exist a zero mode and finite massive bound KK modes. We plot the shapes of the potential for different $q,p$ in Fig.~\ref{FigVfinite}.

For the zero mode, we have
\begin{equation}
  \int U_0^2dz=\int \text{e}^{(p-2q)A}dz\propto\int\text{e}^{-k_2(p-2q)z}dz.
\end{equation}
So the zero mode can be localized on the brane if $q<p/2$. For example, for the case of $p=3$, both the scalar field $(q=0)$ and the vector field $(q=1)$ can be localized on the brane.

While for the massive KK modes, we mainly focus on the effect of the parameters $p$ and $q$ on them. For the scalar field ($q=0$ ), the dimension of the brane $p$ will not affect the effective potential; so the mass spectra of the bound scalar KK modes are the same on different $p-$branes. But for other $q-$form fields, the depth of the effective potential increases with $p$, and so does the number of the massive bound KK modes. For example, for the vector field ($q=1$), the mass spectra are
\begin{eqnarray}
  m^2&=&(0, 6.20)\cup (7.71, \infty), ~~~~~~~~~~~~~~~~~~~\text{for}~~~p=3,\\
  m^2&=&(0, 12.93, 21.49)\cup (25, \infty), ~~~~~~~~~~~~\text{for}~~~p=5,\\
  m^2&=&(0, 15.80, 27.30, 34.05)\cup (35.43, \infty), \text{for}~~~p=7,
\end{eqnarray}
from which we see that the number of the massive bound KK modes increases with the dimension of the $p-$brane.

Moreover, on the same $p-$brane, the depth of the potential decreases with $q$, thus the number of the massive bound KK modes also decreases. For example, on the $7-$brane, there are more massive bound KK modes of the scalar field ($q=0$) than that of the higher form fields, which can be seen from the following mass spectra:
\begin{eqnarray}
  m^2&=&(0, 22.96,41.72,56.04,65.51)\cup (69.44, \infty), ~~~~\text{for}~~~q=0,\\
  m^2&=&(0, 15.80, 27.30, 34.05)\cup (35.43, \infty), ~~~~~~~~~~~~\text{for}~~~q=1,\\
  m^2&=&(0, 8.61)\cup (12.76, \infty), ~~~~~~~~~~~~~~~~~~~~~~~~~~~~~~\text{for}~~~q=2.
\end{eqnarray}

\subsubsection{Pure geometrical thick $p-$brane}

At last we consider the pure geometrical thick $p-$brane. For the $D-$dimensional Weyl-integrable manifold $M_D^W$, which is specified by a pair of $(g_{MN}, \omega)$ with $g_{MN}$ the $D-$dimensinal metric and $\omega$ a Weyl scalar function, the action of the Kaluza-Klein theory is \cite{KeYang2009Weyl,BarbosaCendejas2006,Arias2002,BC2007WeylGravity}
\begin{equation}\label{actionW}
  S_D^W=\int_{M_D^W}d^Dx\sqrt{-g}\;\text{e}^{-\frac{p}{2}\omega}
  \left[\frac{1}{2\kappa_D^2}R-\frac{3}{2}\xi(\nabla\omega)^2-3U(\omega)\right],
\end{equation}
where $\xi$ is the coupling constant and $U(\omega)$ is the self-interaction potential of the Weyl scalar function $\omega$. By performing a conformal transformation $\hat{g}^{MN}=\text{e}^{-\omega}g^{MN}$, the action (\ref{actionW}) is mapped into the Einstein frame:
\begin{eqnarray}\label{actionR}
  S^R_D=\int d^Dx\;\sqrt{-\hat{g}}\left[\frac{1}{2\kappa_D^2}\hat{R}-\frac{3}{2}\xi(\hat{\nabla}\omega)^2
  -3\hat{U}(\omega)\right]
\end{eqnarray}
with $\hat{U}=\text{e}^{\omega}U$. Then we can obtain one solution in a compact manifold along the extra dimension with the range $-\frac{\pi}{2}\leq ky\leq\frac{\pi}{2}$:
\begin{eqnarray}
  \omega&=&-\frac{2c(p+1)}{p}\ln{\cos{(k y})},\\
  A&=&c\;\ln{\cos(k y)},
\end{eqnarray}
where $c=\frac{12p^2}{(p+1)[p-12\xi(p+1)]}$, and $k$ is a positive constant. Here, in order to get a finite warp factor $\text{e}^{2A}$ everywhere, we need $c>0$. Further, in order for the photon to spend infinite time from the brane to the boundary, we require $c\ge 1$. The solution in 5-dimensional space-time can be found in Refs.~\cite{Liu2008WeylPT,BarbosaCendejas2006}.

\begin{figure*}[htb]
\begin{center}
\subfigure[$c>0~(-\pi/2<ky<\pi/2)$]{\label{figVWeyla}
\includegraphics[width=6cm]{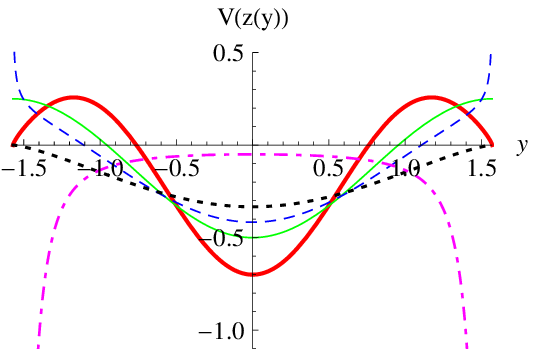}}
\subfigure[$c\ge 1~ (-\infty<z<\infty)$]{\label{figVWeylb}
\includegraphics[width=6cm]{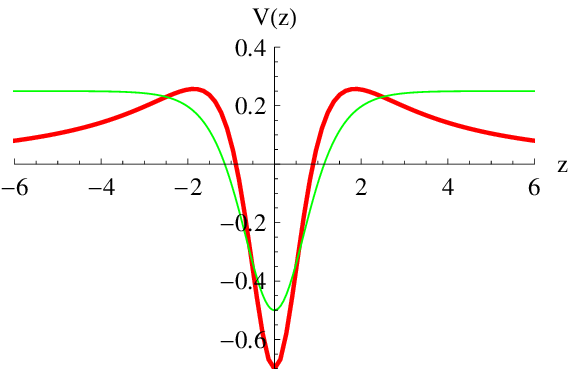}}
\subfigure[$\frac{2}{2+p-2q}{\le}c<1 ~ (-z_c <z<z_c)$]{\label{figVWeylc}
\includegraphics[width=6cm]{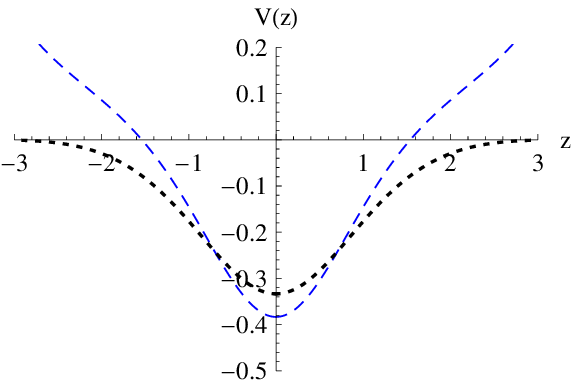}}
\subfigure[$0<c<\frac{2}{2+p-2q} ~ (-z_c <z<z_c)$]{\label{figVWeyld}
\includegraphics[width=6cm]{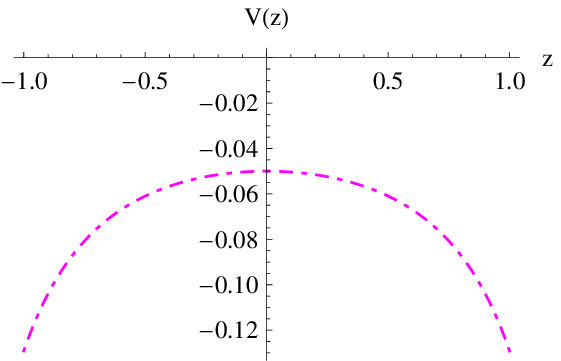}}
\end{center}
 \caption{The shapes of the effective potential $V(z(y))$ and $V(z)$ for pure geometrical $p-$brane with $k=1$, $p=3$, $q=1$, and $c=1.4$ for the thick lines, $c=1$ for the thin solid lines, $c=0.76$ for the dashed lines, $c=0.67$ for the dotted lines, and $c=0.1$ for the dot dashed lines.}
 \label{figVWeyl}
\end{figure*}

Through the coordinate transformation $dz=\text{e}^{-A}dy$, the effective potential (\ref{V}) can be rewritten in $y$-coordinate as
\begin{eqnarray}\label{VW}
  V(z(y))&=&\text{e}^{2A}\left[ \frac{(p-2q)^2+2(p-2q)}{4}(\partial_y A)^2
  +\frac{p-2q}{2}\partial_{y}^2 A\right]\nonumber\\
      &=&\frac{1}{4}c(p-2q)k^2\cos^{2(c-1)}(ky)\big[c(2+p-2q)\sin^2{(ky)}-2\big],
\end{eqnarray}
from which we can see that the potential $V(z(y))$ has different types of shapes decided by the parameter $c$,
and it is the same for $V(z)$. We plot the shapes of the $V(z(y))$ and $V(z)$ for different $c$ in Fig.~\ref{figVWeyl}.
Note that, for $c\ge 1$, the range of the $z$ is infinite; while for $0<c < 1$,
the range of the $z$ is finite, i.e., $-z_c <z<z_c$ with $z_c=\sqrt{\pi}~\Gamma(\frac{1-c}{2})/2\Gamma(1-\frac{c}{2})$.

From (\ref{VW}) it is clear that, for $c>1$ and $q<p/2$, $V(y=0)=-{ck^2}(p-2q)/{2}<0$ and $V(ky\rightarrow\pm\pi/2)\rightarrow 0_{+}$, and the effective potential is a volcano-like one. So there exists a zero mode for $q<p/2$. With the orthonormality condition (\ref{NormCondi})
\begin{eqnarray}\label{zeroMW}
\int_{-z_b}^{z_b}U_0^2dz&=&2\int_{0}^{z_b} \text{e}^{(p-2q)A}dz\nonumber\\
&=&\int_0^{\frac{\pi}{2k}} \text{e}^{(p-2q-1)A}dy
=\int_0^{\frac{\pi}{2k}} \cos^{w}{(ky)} dy\nonumber\\
&=&
  \frac{\sqrt{\pi}~\Gamma(\frac{1+w}{2})}{2\Gamma(1+\frac{w}{2})}
  - \frac{\cos^{1+w}{(\pi/2)}}{1+w},
\end{eqnarray}
where $w=c(p-2q-1)$, it is found that when $q<\frac{p-1+\frac{1}{c}}{2}$ the above integral is finite, hence, the zero mode can be localized on the brane. Considering $c>1$, the final result is that the $q-$form fields with $q<p/2$ have a localized zero mode, which is different from that on the thin $p-$brane and on the one with infinite extra dimension.

In fact, for the zero mode of the $q-$form field, the $\int U_0^2dz$ in the physical coordinate system becomes $\int \text{e}^{(p-2q-1)A}dy$; it is always finite for $q=(p-1)/2$ and finite $y$. So the zero mode of the fields with $q=(p-1)/2$ can always be localized on the brane with finite extra dimension. For the pure geometrical thick brane with compact extra dimension considered here, the $q-$form field with $q=(p-1)/2~(<p/2)$ can also be localized on the brane.

While for the case $c=1$, the potential is a PT-like one for $q<p/2$ (see Fig. \ref{figVWeylb}), whose behavior is $V(y=0)=-k^2(p-2q)/2 <0$ and $V(ky\rightarrow\pm\pi/2)\rightarrow k^2(p-2q)^2/4>0$, so there must exist a zero mode and some massive bound KK modes. From (\ref{zeroMW}), it is clear that the zero mode can be localized for $q<p/2$. And the number of the massive bound KK modes of the $q-$form field increases with the dimension of the $p-$brane. Here, the parameter $p$ also effects the number of the massive bound scalar KK modes, which is different with that on the thick flat $p-$brane with finite extra dimension discussed in section \ref{flatfinite}.

For $\frac{2}{2+p-2q}<c<1$ and $q<p/2$, $V(y=0)=-c(p-2q)k^2/2<0$ and $V(ky\rightarrow\pm\pi/2)\rightarrow\infty_+$, so the effective potential is an infinite potential well (see Fig. \ref{figVWeylc}). Thus, there are a bound zero mode and infinite number of massive bound KK modes, all these KK modes can be localized on the brane.

For $c=\frac{2}{2+p-2q}$ and $q<p/2$, we will get a negative PT-like potential (see Fig. \ref{figVWeylc}), and the zero mode can be localized on the brane, which can be seen from (\ref{zeroMW}).

We do not consider the case of $0<c<\frac{2}{2+p-2q}$, for which the potential approaches negative infinity at the boundary (see Fig. \ref{figVWeyld}).

\section{Conclusion}
\label{conclusion}

In this paper, we first investigated one general method of localizing the $q-$form fields on $p-$branes. We found that the KK modes of the $q-$form fields satisfy the Schr\"{o}dinger-like equation. Thus, the characters of the KK modes are decided by the behavior of the effective potential, which depends on the parameters $q$ and $p$ and the background geometry.

Then we applied this to some typical $p-$brane models. It was found that for the thin Minkowski $p-$brane the effective potential is a volcano-like one, which results in that there exist a zero mode and continuous massive KK modes, and only for $q<(p-1)/2$ the zero mode of the $q-$form field can be localized on the brane. And it is the same for the thick Minkowski $p-$brane with infinite extra dimension. While for the thick Minkowski $p-$brane with finite extra dimension, there is a PT-like potential. So a zero mode and some massive bound KK modes can exist. For the fields with $q<p/2$, the zero mode can be localized on the brane. The number of the massive bound KK modes for the same $q-$form field increases with the dimension of the $p-$brane, except for the scalar field ($q=0$). Moreover, on the same $p-$brane, the $q-$form fields with higher $q$ have less number of the massive bound KK modes.

For the pure geometrical thick $p-$brane, the shape of the effective potential is decided by the value of the parameter $c$. For $c>1$ and $q<p/2$, the effective potential is a volcano-like one, and the condition for localizing the zero mode of the $q-$form field is $q<p/2$, which is different from that on the thin $p-$brane and on the one with infinite extra dimension. For $c=1$ and $q<p/2$, the effective potential is a PT-like one, and only the fields with $q<p/2$ have a localized zero mode. The number of the massive bound KK modes for any $q-$form field, including the scalar field, increases with the dimension of the $p-$brane. And for $\frac{2}{2+p-2q}<c<1$ and $q<p/2$, the effective potential is an infinite potential well, and there exist a localized zero mode and infinite number of massive bound KK modes. For $c=\frac{2}{2+p-2q}$ and $q<p/2$, there is a negative PT-like potential, and the zero mode can also be localized on the brane. However, we only need the case of $c\ge 1$ in order to secure a fully predictive initial value Cauchy problem \cite{Mannheim2005}.

\section*{Acknowledgement}

This work was supported by the Program for New Century Excellent Talents
in University, the National Natural Science Foundation of China (No. 11075065),
the Doctoral Program Foundation of Institutions of Higher Education of China
(No. 20090211110028), the Huo Ying-Dong Education Foundation of
Chinese Ministry of Education (No. 121106), and the Fundamental Research Funds for the Central Universities (No. lzujbky-2012-207).


\end{document}